\documentclass[aip, amsmath,amssymb,reprint]{revtex4-1}

\usepackage{graphicx}% Include figure files
\usepackage{dcolumn}% Align table columns on decimal point
\usepackage{bm}% bold math
\usepackage{xcolor}

\usepackage[utf8]{inputenc}
\usepackage[T1]{fontenc}
\usepackage{mathptmx}
\usepackage{caption}
\usepackage[font=small,labelfont=bf,
   justification=justified,
   format=plain]{caption}

\usepackage{booktabs}
\usepackage{hologo}
\usepackage{array} % required for text wrapping in tables

\begin{document}

\preprint{AIP/123-QED}

\title[]{Simultaneous study of acoustic and optic phonon scattering of electrons and holes in undoped $\mathrm{GaAs}$/$\mathrm{Al_{x} Ga_{1-x} As}$ heterostructures}
% Force line breaks with \\

\author{Y. Ashlea Alava}
% \email{y.ashleaalava@unsw.edu.au}
 %Lines break automatically or can be forced with \\
\author{K. Kumar}%
%\\This line break forced with \textbackslash\textbackslash
%
\affiliation{
School of Physics, University of New South Wales, Sydney, New South Wales 2052, Australia%\\This line break forced with \textbackslash\textbackslash
}%
\affiliation{%
ARC Centre of Excellence for Future Low-Energy Electronics Technologies, University of New South Wales, Sydney, NSW 2052, Australia %\\This line break forced% with \\
}%

\author{C. Harsas}%
\affiliation{
School of Physics, University of New South Wales, Sydney, New South Wales 2052, Australia%\\This line break forced with \textbackslash\textbackslash
}%

\author{P. Mehta}%
\affiliation{
School of Physics, University of New South Wales, Sydney, New South Wales 2052, Australia%\\This line break forced with \textbackslash\textbackslash
}%

\author{P. Hathi}
\affiliation{
School of Physics, University of New South Wales, Sydney, New South Wales 2052, Australia}%\\This line break forced with \textbackslash\textbackslash}

\author{C. Chen}
\author{D. A. Ritchie}
% \homepage{http://www.Second.institution.edu/~Charlie.Author.}
\affiliation{%
Cavendish Laboratory, University of Cambridge, Cambridge, CB3 0HE, United Kingdom%\\This line break forced% with \\
}%

 \author{A. R. Hamilton}
 \email{alex.hamilton@unsw.edu.au}
% \email{Alex.Hamilton@unsw.edu.au}%Lines break automatically or can be forced with \\
\affiliation{
School of Physics, University of New South Wales, Sydney, New South Wales 2052, Australia%\\This line break forced with \textbackslash\textbackslash
}%
\affiliation{%
ARC Centre of Excellence for Future Low-Energy Electronics Technologies, University of New South Wales, Sydney, NSW 2052, Australia %\\This line break forced% with \\
}%

\date{\today}% It is always \today, today,
             %  but any date may be explicitly specified

\begin{abstract}

The study of phonon coupling in doped semiconductors via electrical transport measurements is challenging due to unwanted temperature-induced effects such as dopant ionisation and parallel conduction.
Here, we study phonon scattering in 2D electrons and holes in the $1.6-92.5$K range \textcolor{black}{without the use of extrinsic doping}, where both acoustic and longitudinal optic (LO) phonons come into effect.
We use undoped GaAs/$\mathrm{Al_{x} Ga_{1-x} As}$ heterostructures and examine the temperature dependence of the sample resistivity, extracting phonon coupling constants and the LO activation energy.
Our results are consistent with results obtained through approaches other than transport measurements, and highlight the benefit of this approach for studying electron-phonon and hole-phonon coupling.

\end{abstract}

\maketitle

Understanding the omnipresent scattering of charge carriers and phonons in semiconductors is relevant to a wide range of quantum physics topics; from phonon-assisted cooling to electron hydrodynamics and quantum computing. Phonon coupling determines how efficiently charge carriers can lose energy to the lattice in semiconductor devices, and thus how much current they can sustain before hot electron effects come into play\cite{shah1978hot}, as well as how easy it is to cool electrons to low temperatures\cite{bistritzer2009electronic, tse2009energy, bernard2012probing}. Phonons also play a crucial role in electron hydrodynamics, where they determine the conditions under which hydrodynamic flow can be observed\cite{ho2018theoretical, keser2021geometric, gupta2021hydrodynamic, estrada2024alternative, gusev2018viscous}. Additionally, phonons are well-known to cause decoherence in qubits\cite{camenzind2018hyperfine, camenzind2022hole, huang2024high}, so it would be advantageous to have a simple technique to measure phonon coupling in real devices.
% [\textit{Background}: What have others done?
It can be difficult however to study phonon scattering via electrical transport measurements in doped semiconductor devices because varying the temperature not only changes the phonon scattering but also causes changes in the free carrier density due to thermal excitation of donors, and can lead to an increase in Coulomb scattering from the ionised donors\cite{Lin1984ParallelConduction}. These unwanted effects make it hard to isolate and therefore study the phonon contribution to the conductivity, $\sigma$(T).
\vspace{0.1cm}

Here we introduce a simple technique for studying acoustic and LO phonon coupling in semiconductor heterostructure devices \textcolor{black}{without extrinsic doping and on the same epitaxial stack for N and P-type devices. This ensures that the extrinsic scattering is the same for both carrier types, while also circumventing unwanted dopant excitation effects such as increased scattering and parallel conduction.} We use electrical transport measurements where the sample temperature is varied causing a change in conductivity due to phonon scattering, while the carrier density is kept constant to ensure the variation in $\sigma$(T) is due solely to the change in temperature.
%Benefit of undoped is definitely that you can study holes and electrons
\vspace{0.1cm}

\begin{center}
%\vspace{1cm}
\includegraphics[width=6.2cm]{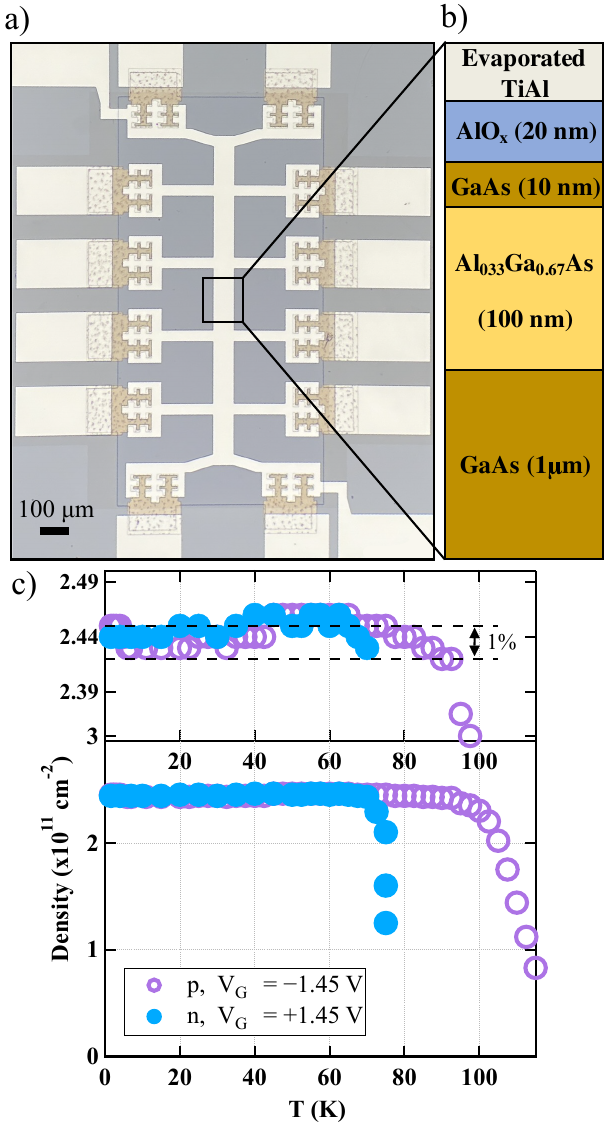}
\captionof{figure}[]{(a) Optical image of the fabricated Hall Bar with an overall top gate. (b) Cross-sectional depiction of the fabricated devices. (c) Carrier density plotted against temperature for a fixed top gate \textcolor{black}{providing a carrier density of $n, p = 2.45 \times 10^{11}$ cm$^{-2}$ for electrons and holes.} The graph shows the carrier density remains constant to better than $1\%$ when increasing the temperature from $1.6$K to $72.5$K for electrons (solid, blue trace) and to $92.5$K for holes (hollow, purple trace), above which the carrier density rapidly decreases with temperature and time.
}
\label{FIG1}
\end{center}

\begin{figure*}
\vspace{1cm}
\includegraphics[width=18cm, height=9cm]{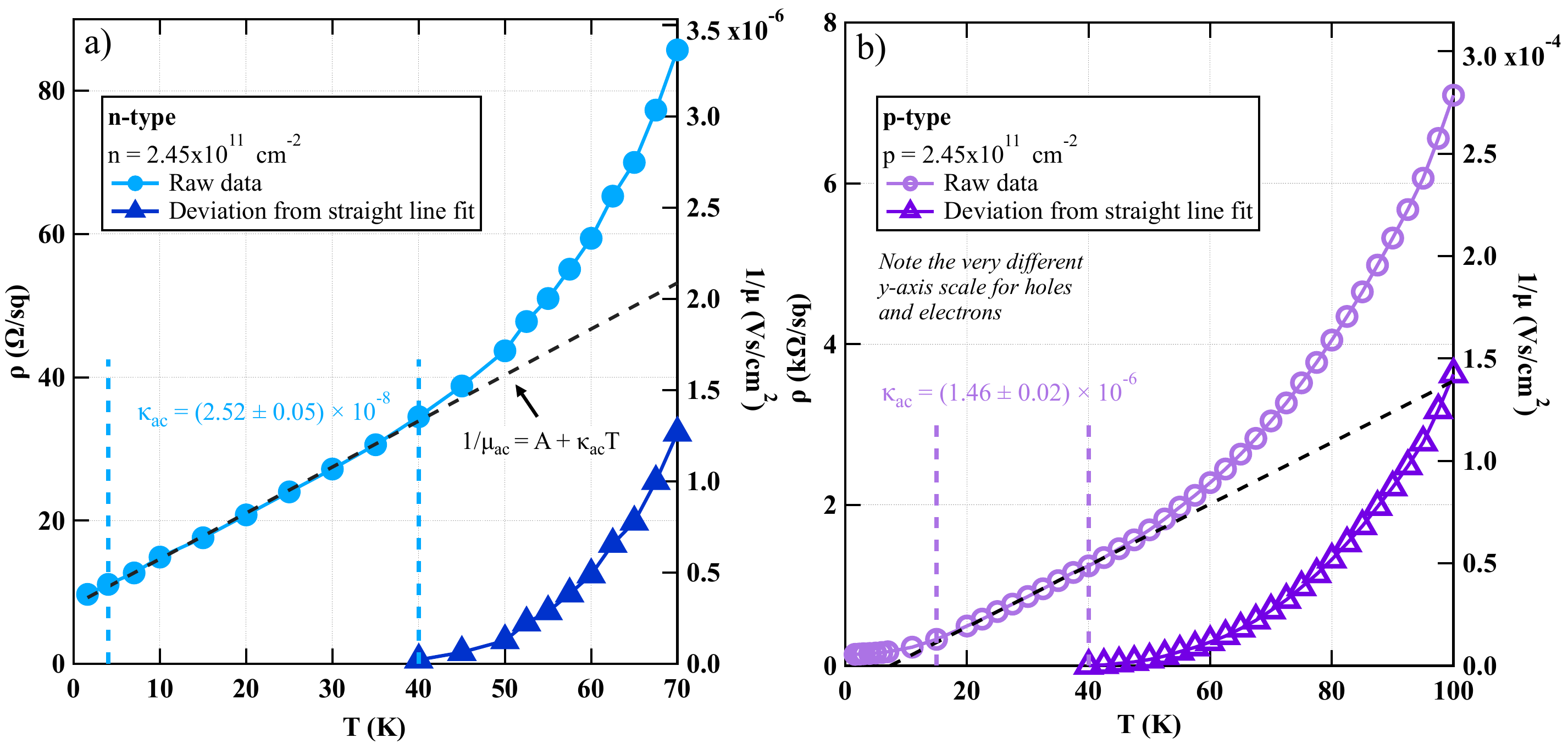}
\captionof{figure}[]{(a) Resistivity of the N-type device at $n=2.45\times 10^{11}$ cm$^{-2}$: $\rho$ (left axis) and $1/\mu$ (right axis) plotted against temperature. The light blue circles are the raw data and the dashed line is a straight line fit to the acoustic phonon region $10-40$K with slope $\kappa_{ac}=(2.52\pm 0.05) \times 10^{-8}$ Vs/Kcm$^2$ (acoustic coupling constant). Subtracting the straight line fit from the raw data yields the LO phonon trace (dark blue triangles). (b) Analogous plot for the P-type device (note the difference in y-axis range between electrons and holes) with an extracted acoustic phonon coupling constant: $\kappa_{ac}=(1.46\pm 0.02) \times 10^{-6}$ Vs/Kcm$^2$, for $p=2.45\times 10^{11}$ cm$^{-2}$.}
\label{FIG2}
\end{figure*}
\vspace{0.5cm}

\noindent This technique is generic; here we demonstrate it with an undoped GaAs/$\mathrm{Al_{0.33} Ga_{0.67} As}$ heterostructure in accumulation mode where the carrier density remains constant to within $1\%$ from $1.6$K up to $72.5$K for electrons and $92.5$K for holes.

Accumulation mode (N and P-type) MOSFET Hall Bars of width $70$ $\mu$m and length $1150$ $\mu$m were investigated (see Figure \ref{FIG1}a). The wafer used in this study (W1789) was grown by MBE on a (100) GaAs substrate. The growth sequence consisted of a substrate degas, followed  by a $1\mu$m GaAs buffer, 100 nm of undoped 33\% $\mathrm{Al_{0.33} Ga_{0.67} As}$ , and a 10 nm GaAs cap (see Figure \ref{FIG1}b).

To make ohmic contact to the wafer, a pit was etched \textcolor{black}{through the semiconductor heterostructure $50$ nm past the 2DEG plane into the GaAs buffer layer, where the 2DEG is formed}. Then, ohmic metal was evaporated; NiAuGe for the N-type device, and AuBe for the P-type device. The devices were then annealed for $90$ seconds at $450^{\circ}$C and $490^{\circ}$C, respectively.\textcolor{black}{\footnote{\textcolor{black}{Note that the only difference in fabrication process between the N and P-type devices is the ohmic material used and the anneal temperature.}}} A $20$ nm layer of aluminium oxide (AlO$_x$) was grown at $200^{\circ}$C by atomic layer deposition (ALD) on the wafer surface to insulate the ohmic contacts and the subsequently evaporated TiAl top gate. The carrier density (n, p) was measured by sweeping a small magnetic field ($\sim 200$ mT) perpendicular to the 2D carrier plane and measuring the Hall voltage.
The N-type device shows constant electron density to within $1\%$ for a fixed gate voltage $V_G=1.45$V up to $72.5$K (solid, blue symbols, Figure \ref{FIG1}c) and the P-type device shows constant hole density to within $1\%$  for $V_G=\textcolor{black}{-}1.45$V up to $92.5$K (open purple symbols, Figure \ref{FIG1}c). The carrier density drops dramatically past $72.5$K ($92.5$K), likely due to unwanted thermally-activated charge accumulation at the semiconductor-AlO$_x$ interface, which causes screening of the gate electric field\cite{AshleaAlavaAPL, ashlea2022ultra}.
\vspace{0.1cm}

To study the phonon coupling, we fix the carrier density ($n, p = 2.45 \times 10^{11}$ cm$^{-2}$) and gradually increase the sample temperature to measure the change in sample resistivity $\rho (T)=1/\sigma (T)$. \textcolor{black}{The sample resistivity is measured by applying a source-drain voltage $V_{SD}=100$ $\mu$V, which produces source drain currents below $50$nA, minimising sample heating}.
Figure \ref{FIG2}a shows a linear dependence of $\rho$ with temperature (left axis) for the N-type device in the $\textcolor{black}{4}-40$K temperature range. This is due to acoustic phonon scattering, and is consistent with previous findings\cite{kawamura1992phonon, Kawamura1990, Foxon_1989, HARRIS1990113, hwang2008limit}. Above $40$K, the resistivity starts to increase rapidly due to LO phonon scattering\cite{basu1980lattice, price1981two, walukiewicz1984electron,arora1985phonon}. The right hand side axis of Figure \ref{FIG2}a is the inverse mobility $1/\mu=\rho n e$, i.e, the resistivity scaled by the density. \textcolor{black}{The total inverse mobility can be written as:}

\begin{equation}
    \textcolor{black}{1/\mu (T) = 1/\mu_{0} + 1/\mu_{ac}(T)+
    1/\mu_{opt}(T)},
\end{equation}
\textcolor{black}{where $1/\mu_{0}$ is the zero-temperature inverse mobility, and $1/\mu_{ac}$ and $1/\mu_{opt}$ are the acoustic and LO optic phonon components to the total mobility, respectively. The zero-temperature inverse mobility is dominated by Coulomb scattering from background impurities, remote ionised impurities and interface roughness effects; relevant below $4$K\cite{ashlea2022ultra, AshleaAlavaAPL}. Acoustic phonons dominate in the $4-40$K temperature range, and thus a} straight line fit to $1/\mu$ vs T in this range enables the extraction of the acoustic phonon scattering rate\cite{mendez1984temperature}:
\begin{equation}
    1/\mu_{ac}(T)=\kappa_{ac} T+A
\end{equation}
with $\kappa_{ac}=(2.52\pm 0.05) \times 10^{-8}$ Vs/Kcm$^2$ for electrons, where the uncertainty comes from the linear regression. These values agree to within 1\% with data from Ref. \onlinecite{walukiewicz1988acoustic}.
\textcolor{black}{For holes, the experimental data (hollow circles, Figure \ref{FIG2}b) deviate from a straight line below $15$K, which may be due to screening of the deformation potential and entering the Bloch-Gruneisen regime\cite{simmons1994optimization}. When fitting a straight line through the linear region $15-40$K\textcolor{black}{\footnote{We thank the referee for highlighting the apparent negative intercept of resistivity from the fitting of the linear slope for holes, which is also observed in Ref. \onlinecite{walukiewicz1988acoustic}. This may indicate that screening of the deformation potential is significant for holes at low temperatures.}}, we obtain a value for the acoustic coupling constant of}
approximately $60$ times higher than for electrons $\kappa_{ac}=(1.46\pm 0.02) \times 10^{-6}$ Vs/Kcm$^2$. \textcolor{black}{The value we obtain is within $9$\% of the value reported in Ref. \onlinecite{walukiewicz1988acoustic}}. This small difference may be due to complexities of the hole bandstructure, which is very sensitive to the 2D confining potential\cite{eisenstein1984effect, habib2009spin}.

\begin{center}
%\vspace{1cm}
\includegraphics[width=1\linewidth]{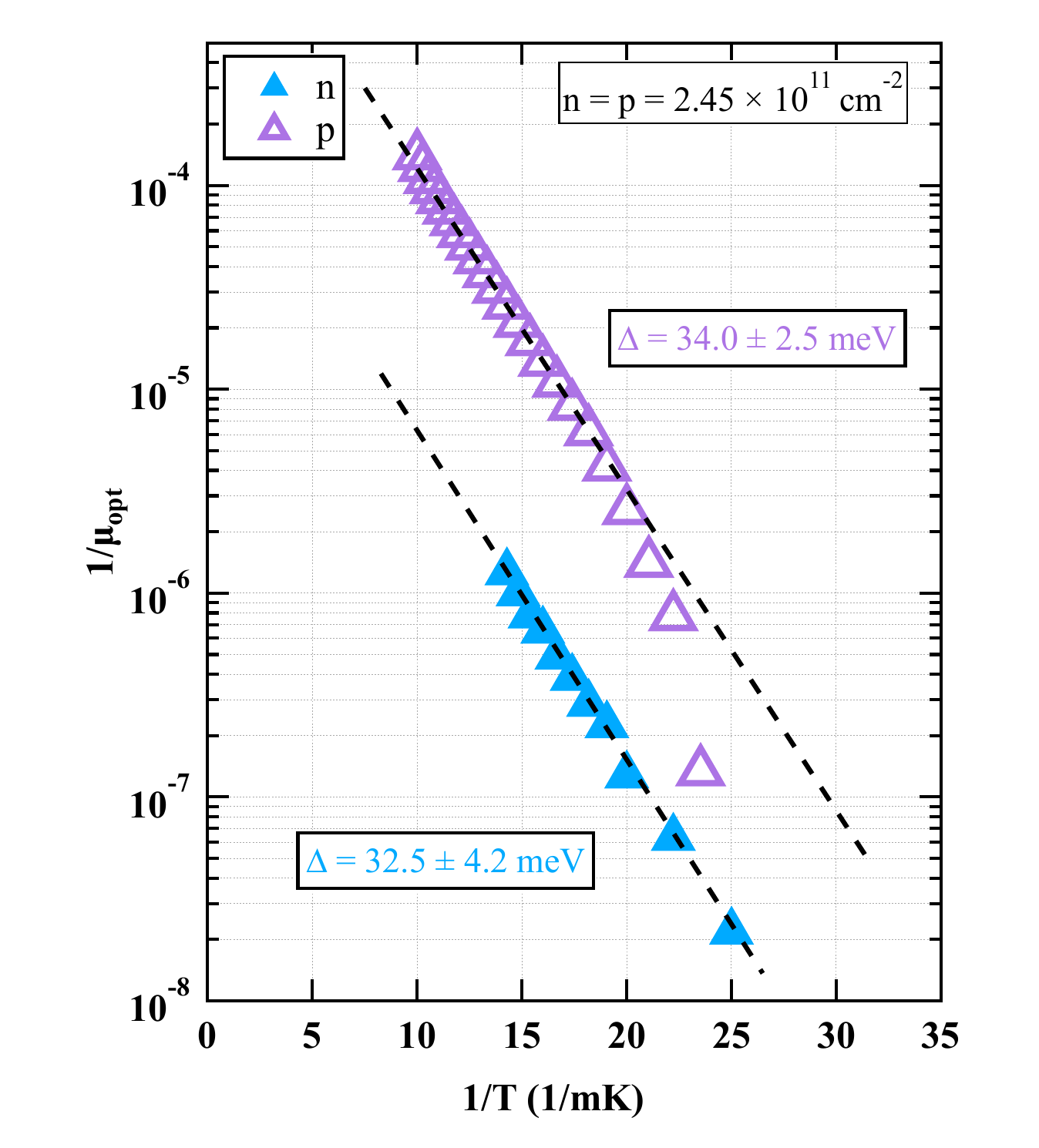}
\captionof{figure}[]{Semi-log plot of the inverse mobility in the LO phonon regime $1/\mu_{opt}$ as a function of $1/T$ for electrons (solid, blue trace) and holes (hollow, purple trace). The dashed lines are straight line fits from which the LO phonon activation energy $\Delta$ is extracted.
}
\label{FIG3}
\end{center}
The large difference in the acoustic phonon coupling for holes and electrons ($\kappa_{ac}^{h}/\kappa_{ac}^{e} \approx 57$), is likely due to the difference in effective mass. Following calculations from Ref. \onlinecite{walukiewicz1988acoustic}, the acoustic coupling constant ratio between holes and electrons is proportional to the ratio of hole and electron mass $\kappa_{ac}^{h}/\kappa_{ac}^{e} \sim (m_{h}^{*}/m_{e}^{*})^{7/3}$. We can estimate the hole mass $m_{h}^{*}$ by using the extracted acoustic phonon constants from Figure \ref{FIG2} and taking $m_e^*=0.067$m$_e$, which gives a hole mass of $0.38$m$_e$, where m$_e$ is the electron mass in free space. This value is close to that obtained from other transport measurements of $(100)$ hole systems, and somewhat larger than the band-edge mass of $0.2m_e$ due to the $k^3$ Rashba effect in holes\cite{nichele2014spin}.

Unlike acoustic phonons, LO phonons have a minimum activation energy\cite{vinter1984phonon} so are thermally activated. The LO phonon contribution to the inverse mobility $1/\mu_{opt}$ follows a Boltzmann distribution: %\cite{menzinger1969meaning}:

\begin{equation}
    1/\mu_{opt}(T)=\kappa_{opt}e^{-\Delta/k_B T}
    \label{eqOptiPhonons}
\end{equation}

\noindent where $\kappa_{opt}$ is the LO phonon coupling strength, $\Delta$ is the optic phonon energy, and $k_B$ the Boltzmann constant. \noindent \textcolor{black}{The LO phonon component of the inverse mobility (triangles in Figure \ref{FIG2}a, b) can be extracted and isolated by subtracting the acoustic phonon component from the raw data, since thermally activated LO phonons only contribute to the scattering for $T>40$K.}

\begin{table}
    \centering
    \footnotesize
\caption{Comparison table of acoustic and LO phonon coupling constants with values from the literature}
\label{tab:my_label}
    \begin{tabular}{>{\raggedright\arraybackslash}p{0.10\linewidth}>{\centering\arraybackslash}p{0.18\linewidth}>{\centering\arraybackslash}p{0.19\linewidth}>{\centering\arraybackslash}p{0.18\linewidth}>{\centering\arraybackslash}p{0.18\linewidth}>{\centering\arraybackslash}p{0.09\linewidth}}
    \toprule
     \textbf{ Device} & \textbf{Measurement type} & \textbf{Density} ($\times10^{11}$ cm$^{-2}$) & \bm{$\kappa_{ac}$} (Vs/Kcm$^{-2}$) & \bm{$\kappa_{opt}$} (Vs/cm$^{-2}$) & \bm{$\Delta$} (meV)\\
\midrule
       Wafer W1789 &  Transport (n-type)&  2.45&  (2.52 $\pm$ 0.05) $\times 10^{-8}$&  (2.66 $\pm$ 0.24) $\times 10^{-4}$& 32.5 $\pm$ 4.2\\
 & (p-type)& 2.45& (1.46 $\pm$ 0.02) $\times 10^{-6}$& (5.63 $\pm$ 0.11) $\times 10^{-3}$&34.0 $\pm$ 2.5\\
        Ref. \onlinecite{keser2021geometric}&  Transport (n-type)&  2.45&  2.55$\times 10^{-8}$&  & \\
         Ref. \onlinecite{mendez1984temperature}&  Transport (n-type)&  $\sim$ 2.5&  $\sim$ 2.7$\times 10^{-8}$&  & \\
         Ref. \onlinecite{walukiewicz1988acoustic}&  Transport (n-type)&  1.6&  2.60$\times 10^{-8}$&  & \\
 & (p-type)& 2.7& 1.6$\times 10^{-6}$& &\\
         Ref. \onlinecite{langerak1988carrier}&  Resonant polaron cyclotron resonance&  &  &  & 36.7\\
         Ref. \onlinecite{dzurak1992two}&  Hot electron injection&  &  &  & 36\\
         Ref. \onlinecite{lockwood2005optical}&  Infrared spectroscopy&  &  &  & 36.2\\
         \bottomrule
    \end{tabular}

\end{table}

Plotting $1/\mu_{opt}$ vs 1/T (see Figure \ref{FIG3}) on semilog axes allows the LO phonon energy $\Delta$ to be extracted from a straight line fit (dashed lines Figure \ref{FIG3}). This yields almost identical values for electrons and holes; $\Delta = 32.5 \pm 4.2$ meV and $\Delta = 34.0 \pm 2.5$ meV\footnote{The uncertainty in $\Delta$ comes from the subtraction of the acoustic phonon component (straight line fit, Figure \ref{FIG2}a, b) to the raw data (light colour circles in Figure \ref{FIG2}a, b). Using the uncertainty ($\Delta \kappa_{ac}$) in the slope ($\kappa_{ac}$) of the straight line fit in Figure \ref{FIG2}, one can obtain two worst possible fits, with slopes $\kappa_{ac}+\Delta \kappa_{ac}$ and $\kappa_{ac}-\Delta \kappa_{ac}$. Subtracting these fits to the raw data produces two $1/\mu_{opt}$ vs. 1/T traces, which yield a maximum ($\Delta_{max}$) and a minimum LO phonon activation energy ($\Delta_{min}$). The uncertainty in the activation energy $\Delta$ is then $(\Delta_{max}-\Delta_{min})/2$, which is calculated separately for electrons and holes.}, as expected, since the phonon activation energy is independent of carrier type. On the other hand, the LO phonon coupling constant does depend on carrier type, with a higher value for holes $\kappa_{opt}=(5.63\pm 0.11)\times 10^{-3}$ Vs/cm$^2$ than for electrons $(2.66\pm 0.24)\times 10^{-4}$ Vs/cm$^2$, as observed for the acoustic phonon coupling constant. \textcolor{black}{The ratio of LO phonon coupling constants for holes and electrons is $\kappa_{opt}^{h}/\kappa_{opt}^{e} \approx 21$, which is lower than the ratio of acoustic coupling constants. We currently do not understand the reason for this difference.}
\vspace{0.1cm}

In table \ref{tab:my_label} we compare the phonon coupling constants and activation energy from this work with previous data from the literature. Our results for the acoustic phonon coupling agree well with those obtained in references \onlinecite{keser2021geometric, mendez1984temperature,walukiewicz1988acoustic} for a similar density range\textcolor{black}{\footnote{\textcolor{black}{We extracted the acoustic coupling constant for different densities and observed a weak dependence of the acoustic phonon coupling constant, almost within the error bars. Further future systematic study using this technique at different densities in both N and P-type devices would be needed to conclusively discern the density dependence (or lack thereof)}}}. Similarly, the LO phonon activation energy extracted for the N and P-type samples agrees within error bars to the values from references \onlinecite{langerak1988carrier, dzurak1992two,lockwood2005optical} that used other methodologies.

\vspace{0.1cm}
%Very hyper basic draft that is probably missing key points, but I just want to have something written.
In conclusion, we demonstrate a simultaneous study of acoustic and LO phonons using standard electrical transport measurements, both in electron and hole devices. This technique is generalisable, in principle, to all undoped semiconductor heterostructures.

\begin{acknowledgements}
\noindent
We would like to acknowledge Oleh Klochan, Daisy Wang, Abhay Gupta and Joseph Hillier for experimental assistance, and Adam Micolich for the loan of equipment used in the experiments.
Device fabrication was carried out in part at the Australian National Fabrication Facility (ANFF) at the UNSW node. This work was funded by the ARC Centre of Excellence for Future Low Energy Electronics Technologies (CE170100039), as well as ARC IL23010072 and EP/R029075/1 Non-Ergodic Quantum Manipulation, UK.
\end{acknowledgements}

\section*{Data Availability}

\noindent The data that support the findings of this study are available from the corresponding author upon reasonable request.

\nocite{*}
\bibliography{References}% Produces the bibliography via BibTeX.

\end{document}